\documentstyle[12pt,epsf]{article}

\renewcommand{\bar}[1]{\overline{#1}}

\newcommand{\gege}{\gamma \, e \to \gamma \, e}

\begin{document}
\begin{flushright}
SLAC--PUB--8197\\
July 1999
\end{flushright}
\bigskip\bigskip

\def\grtsim{\,\,\rlap{\raise 3pt\hbox{$>$}}{\lower 3pt\hbox{$\sim$}}\,\,}
\def\lsim{\,\,\rlap{\raise 3pt\hbox{$<$}}{\lower 3pt\hbox{$\sim$}}\,\,}

\thispagestyle{empty}
\flushbottom

\centerline{\Large\bf Compton Scattering at the NLC and Large Extra Dimensions\footnote
{\baselineskip=14pt
Work supported by the Department of Energy, contract DE--AC03--76SF00515.}}
\vspace{22pt}

\centerline{\bf Hooman Davoudiasl}
\vspace{8pt}
  \centerline{\it Stanford Linear Accelerator Center}
  \centerline{\it Stanford University, Stanford, California 94309}
  \centerline{e-mail: hooman@slac.stanford.edu}
\vspace*{0.9cm}

\begin{abstract}

We study Compton scattering, $\gege$, in the context of the recent proposal for Weak Scale Quantum Gravity (WSQG) with large extra
dimensions.  It is shown that, with an ultraviolet cutoff $M_S \sim 1$ TeV for the effective gravity theory, the cross section for this
process at the Next Linear Collider (NLC) deviates from the prediction of the Standard Model significantly.  Our results suggest that, for
typical proposed NLC energies and luminosities, WSQG can be tested in the range 4 TeV$\lsim M_S \lsim$ 16 TeV, making $\gege$ an
important test channel.  

\end{abstract}

\newpage

\section{Introduction} 

It has recently been proposed that the fundamental scale of quantum gravity $M_F$ can be of the order of the weak scale $\Lambda_w$
if we assume that there are $n$ large spatial extra dimensions \footnote{For earlier related work see Ref.  \cite{IA}.}\cite{ADD}. 
Gravitational data at macroscopic scales demand $n \geq 2$ \footnote{The authors of Ref. \cite{RaSu} propose a scenario in which $n =
1$ can be allowed.} and cosmological and astrophysical observations \cite{SNGAM} suggest that $M_F \grtsim 100$ TeV for $n = 2$. 
However, the data from collider experiments at present energies, as well as cosmological and astrophysical considerations for $n \geq
3$, only yield $M_F \grtsim 1$ TeV.  Once larger center of mass energies become available at future experimental facilities, the
predictions of Weak Scale Quantum Gravity (WSQG) can be tested in the TeV regime, as demonstrated by various recent works on this
subject \cite{Recent}.  

In this paper, we consider the possibility of testing WSQG at a future Next Linear Collider (NLC), using TeV-scale Compton scattering
$\gege$.  It has been shown that a high energy and luminosity $e^+ e^-$ collider can yield $\gamma$-beams comparable in energy and
luminosity, using backward Compton scattering of laser photons from the $e^\pm$ beams \cite{Ginzburg}.  We will show that given the
proposed energies and luminosities of the NLC \cite{NLC}, and assuming the above $\gamma$-beams can be obtained, the process
$\gege$ places strong bounds on the effective mass scale of WSQG.

The present work is organized as follows.  In section 2, we present the tree level Standard Model (SM) and WSQG amplitudes for
$\gege$ and the formulas for calculating the relevant cross sections at the NLC.  The results of our computations are given in
section 3.  Section 4 contains our concluding remarks.  Finally, some of the formulas used in our calculations are given in the
appendices.

\section{Amplitudes and Cross Sections}

Here, we present the tree level SM and WSQG amplitudes for the process $\gamma (k_1) \, e (p_1) \to \gamma (k_2) \, e (p_2)$,
where $(k_1, p_1)$ are the 4-momenta of the initial state photon and electron and $(k_2, p_2)$ are the 4-momenta of the final
state photon and electron, respectively.  For the rest of this work, it is assumed that the fundamental scale of gravity $M_F
\grtsim 1$ TeV and that there are $n \geq 2$ compact extra dimensions of size $R$, even though there are astrophysical and
cosmological considerations \cite{SNGAM} that suggest $M_F \grtsim 100$ TeV for $n = 2$.  Then, Gauss' law yields the
relation\cite{ADD}
\begin{equation}
M_P^2 \sim M_F^{n + 2} R^n, 
\label{MP} 
\end{equation}
where $M_P \sim 10^{19}$ GeV is the Planck mass.  The exact relation among $M_P$, $M_F$, and $R$ is presented in the
appendix and depends on the convention and the compactification manifold used.  We will use the effective Lagrangian and
the Feynman rules of Ref. \cite{Han}.  

Let $s \equiv (k_1 + p_1)^2$, $t \equiv (k_1 - k_2)^2$, and $u \equiv (k_1 - p_2)^2$, where $k_1 + p_1 = k_2 + p_2$.  The SM
contributes through the $s$ and $u$ channels with the amplitudes
\begin{equation}
{\cal M}_{_{SM}}^{(s)} = - \left(\frac{4 \, \pi \, \alpha}{s}\right) \varepsilon^*_\mu (k_2) \, \varepsilon_\nu (k_1) \, {\bar u} (p_2) 
\gamma^\mu (2 p_1^\nu + \not\! k_1 \gamma^\nu) u(p_1)
\label{Ms} 
\end{equation}
and 
\begin{equation}
{\cal M}_{_{SM}}^{(u)} = - \left(\frac{4 \, \pi \, \alpha}{u}\right) \varepsilon^*_\mu (k_2) \, \varepsilon_\nu (k_1) \, {\bar u} (p_2) 
\gamma^\nu (2 p_1^\mu - \not\! k_2 \gamma^\mu) u(p_1),
\label{Mu} 
\end{equation} 
respectively, where $\alpha = 1/137$.  The 4-vector $\varepsilon^\mu (k)$ denotes the polarization vector for a photon of 
4-momentum $k$, and $u (p)$ denotes the Dirac spinor for an electron of 4-momentum $p$.

The $t$-channel contribution of WSQG has the amplitude 
\[
{\cal M}_{_{WSQG}}^{(t)} = - \left(\frac{\pi}{M_S^4}\right) D_n (t) \,  
{\bar u} (p_2) [\gamma_\mu (p_{1 \nu} + p_{2 \nu}) + 
\gamma_\nu (p_{1 \mu} + p_{2 \mu})] u(p_1)
\] 
\begin{equation}
\times [(k_1 \cdot k_2) \, C^{\mu \nu, \alpha \beta} + D^{\mu \nu, \alpha \beta}(k_1, k_2)] 
\varepsilon^*_\beta (k_2) \, \varepsilon_\alpha (k_1),
\label{Mt} 
\end{equation}
where $M_S$ is a momentum cutoff for the effective WSQG Lagrangian; we have taken $M_S = M_F$ here.  The function $D_n (x)$ 
depends on $n$ and is given by \cite{Han}
\[
D_n (x) \approx \ln \left(\frac{M_S^2}{|x|}\right) \, \, \, \,  {\rm for} \, \, \, \, n = 2 
\]
and 
\begin{equation}
D_n (x) \approx \left(\frac{2}{n - 2}\right) \, \, \, \, {\rm for} \, \, \, \, n > 2.
\label{Dnx}
\end{equation}
The expressions for $C^{\mu \nu, \lambda \sigma}$ and $D^{\mu \nu, \lambda \sigma} (k, p)$ are presented in Appendix A.  Note
that the expression for ${\cal M}_{_{WSQG}}^{(t)}$ depends on the cutoff scale $M_S \gg s, |t|, |u|$, 
introduced to regulate the divergent sum over the
infinite tower of Kaluza-Klein states.  The form of this dependence is a result of our assumption that $M_S = M_F$.  However, if $M_S$ is taken
to be much smaller than $M_F$ then 
\begin{equation} 
D_n (x) \to \left(\frac{M_S}{M_F}\right)^{(n + 2)} D_n (x) \, \, \, \, {\rm for} \, \, \, \, n \geq 2, 
\label{DMF} 
\end{equation} 
resulting in a suppression \cite{Giudice}.

The total tree level amplitude ${\cal M}^{(TOT)}$ for $\gege$, including the contributions of both the SM and WSQG, is given by
\begin{equation}
{\cal M}^{(TOT)} = {\cal M}_{_{SM}}^{(s)} + {\cal M}_{_{SM}}^{(u)} + w {\cal M}_{_{WSQG}}^{(t)},
\label{MTOT} 
\end{equation}
where $w$ is an arbitrary constant reflecting our lack of knowledge of the fundamental theory of gravity, and hence the sign and
magnitude of the lowest dimension contribution of the effective WSQG Lagrangian to Eq. (\ref{MTOT}).  However, as long as one is only
interested in an order of magnitude estimate of the size of the WSQG contribution, using ${\cal M}^{(TOT)}$ with $w = \pm 1$, as we do
later, is reasonable.  

As mentioned before, high energy and luminosity $\gamma$-beams can be achieved at the NLC, through backward Compton
scattering of laser photons from the high energy $e^\pm$ beams \cite{Ginzburg}.  The $\gamma$ beams that are
obtained in this way have distributions in energy and helicity that are functions of the $\gamma$ energy and the initial polarizations
of the electron beams and the laser beams.  Laser beam polarization $P_l$ can be achieved close to $100\%$, however, electron
beam polarization $P_e$ is at the $90\%$ level.  We take $|P_l| = 1$ and $|P_e| = 0.9$ for our calculations.  

Let $E_e$ be the electron beam energy, and $E_\gamma$ be the scattered $\gamma$ energy in the laboratory frame.  The fraction of
the beam energy taken away by the photon is then 
\begin{equation}
x = \frac{E_\gamma}{E_e}.
\label{x}
\end{equation}
We take the laser photons to have energy $E_l$.  Then, the maximum value of $x$ is given by 
\begin{equation}
x_{max} = \frac{z}{1 + z},
\label{xmax}
\end{equation}
where $z = 4 E_e E_l/m_e^2$, and $m_e$ is the electron mass.  One cannot increase $x_{max}$ simply by increasing $E_l$, since this
makes the process less efficient because of $e^+ e^-$ pair production through the interactions of the laser photons and the
backward scattered $\gamma$ beam.  The optimal value for $z$ is given by
\begin{equation}
z_{_{OPT}} = 2 \left(1 + {\sqrt 2}\right).
\label{zOPT}
\end{equation}
The photon number density $f(x, P_e, P_l)$ and average helicity $\xi_2 (x, P_e, P_l)$ are functions of $x$, $P_e$, $P_l$, and $z$,
however, we always set $z = z_{_{OPT}}$ in our calculations.  We give the expressions for these two functions in Appendix B.  

Let ${\cal M}_{i j k l}$, $i, j, k, l = \pm$, denote the helicity amplitudes for $\gege$, where $(i, j)$ are the helicities of the initial state
$(\gamma, e)$, and $(k, l)$ are the helicities of the final state $(\gamma, e)$, respectively.  We define $|{\cal M}_{i j}|^2$ by 
\begin{equation}
|{\cal M}_{i j}|^2 \equiv \sum_{k, l} |{\cal M}_{i j k l}|^2, 
\label{Mij2} 
\end{equation} 
where the summation is performed over the final state helicities.  We find,
\begin{equation}
|{\cal M}^{(TOT)}_{+ j}|^2 = \frac{- 32 \pi^2}{s \, u} \left[\alpha + w \left(\frac{s \, u \, D_n}{2 \, M_S^4}\right)\right]^2 
\left[s^2 (1 + j) + u^2 (1 - j)\right],
\label{M+j2}
\end{equation}
where $D_n$ is given by Eq. (\ref{Dnx}).  

For various choices of $(P_{e_1}, P_{l_1})$ of the $\gamma$ beam and
$P_{e_2}$ of the electron beam, the differential cross section $d \sigma/d \Omega$ is given by
\begin{equation}
\frac{d \sigma}{d \Omega} = \frac{1}{(8 \, \pi)^2} \int\frac{d x f(x)}{x \, s_{ee}}
\left[\left(\frac{1 + P_{e_2}\, \xi_2(x)}{2}\right)|{\cal M}_{++}|^2 + \left(\frac{1 -   P_{e_2} \, \xi_2(x)}{2}\right)|{\cal M}_{+-}|^2\right],
\label{diffcs}
\end{equation}
where $s_{ee} = 4 E_e^2$.  Different choices of $(P_{e_1}, P_{l_1})$, in $(f(x), \xi_2(x))$, and $P_{e_2}$  yield different polarization cross
sections.  We note that the expressions for $|{\cal M}_{++}|^2$ and $|{\cal M}_{+-}|^2$ are actually functions of the $\gamma \, e$ center of
mass energy squared ${\hat s} = x \, s_{ee}$, and the center of mass scattering angle $\theta_{cm}$.  We also have $t \to {\hat t}$ and $u \to
{\hat u}$, where ${\hat t} = - ({\hat s}/2 ) (1 - \cos \theta_{cm})$ and ${\hat u} = - ({\hat s}/2 ) (1 + \cos \theta_{cm})$.  In the 
following calculations, we use Eq. (\ref{diffcs}) and the cuts    
\begin{equation}
\theta_{cm} \in [\pi/6, 5 \pi/6] \, \, ; \, \, x \in [0.1, x_{max}] 
\label{cuts}
\end{equation}
to obtain the cross sections.

\section{Results}

In this section, we present our numerical results for the expected size of the WSQG effects at the NLC.  However, before discussing the
results, we would like to make a few remarks regarding our calculations.  First of all, as mentioned before, we have assumed $M_S =
M_F$ in our calculations.  The effects of departure from this assumption are given in Eq.  (\ref{DMF}).  Secondly, the only dependence on
the number of extra dimensions $n$ in our computations comes from Eq. (\ref{Dnx}).  We only distinguish between the cases with $n =
2$ and $n > 2$.  In the case with $n = 2$, in the limit $M_S^2 \gg s$, the WSQG amplitude is enhanced logarithmically compared to the
case with $n > 2$.  In our computations, for $n > 2$, we have $\ln (M_S^2/{\hat t}) > 2/(n - 2)$ over most of the parameter space
considered.  We choose $n = 4$ as a representative value for $n > 2$; other choices result in a rescaling of the effective value of $M_S$.  

We present various cross sections with definite polarizations $(P_{e_1}, P_{l_1}, P_{e_2})$ and also the unpolarized cross sections,
obtained by setting $P_{e_1} = P_{l_1} = P_{e_2} = 0$.  Cases with $w = \pm 1$ will also be considered.  We will see that the unpolarized
gravity cross sections are smaller than the corresponding cross sections for a particular optimal polarization, which will turn out to be
$(+, -, +)$.  This is because the $(+, -, +)$ back-scattered $\gamma$-beam has a larger number of hard photons than the
unpolarized beam \cite{Ginz2}.  Our results also show that the cross sections for $w = - 1$ are larger than the ones for $w = +1$, as
evident from Eq.  (\ref{M+j2}).  We will present some of our results using the choice $w = - 1$, to demonstrate the effect of different
signs of $w$ on the results.  However, we note that it is more conservative to choose $w = + 1$, in order to avoid an overestimate of the
effects, and in any case, this is the choice that follows from a straightforward use of the low energy effective Lagrangian.  In the
following, we will present results indicating that the discovery reach of the NLC for the value of the parameter $M_S$ is approximately
the same for $w = \pm 1$.  

\begin{figure}[htbp] 
\centerline{\epsfxsize=10truecm \epsfbox{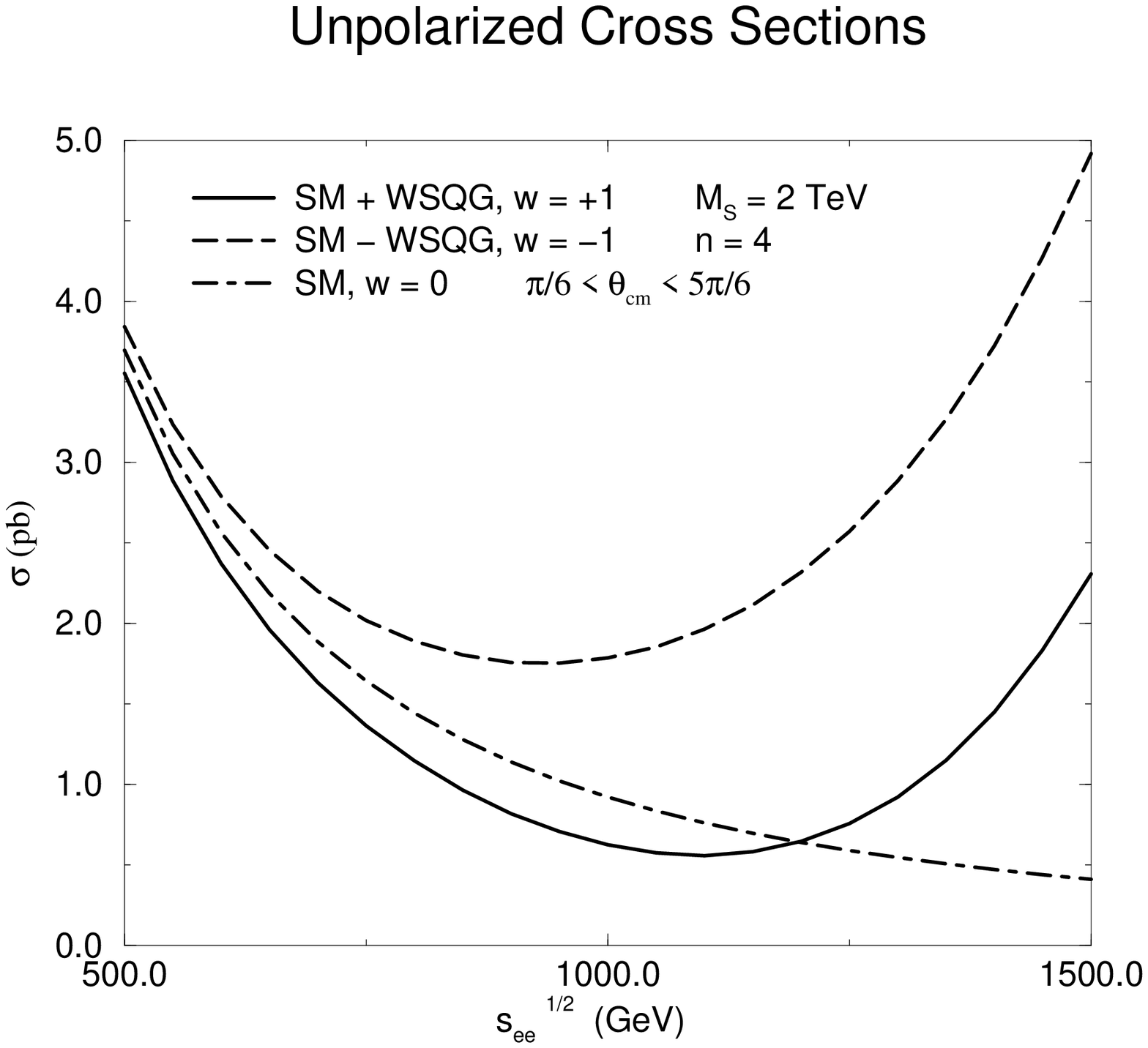}} 
\caption[1]{The unpolarized SM cross section and SM $\pm$ WSQG cross sections with $M_S = 2$ TeV and $n = 4$.} 
\label{unpcs} 
\end{figure}

\begin{figure}[htbp] 
\centerline{\epsfxsize=10truecm \epsfbox{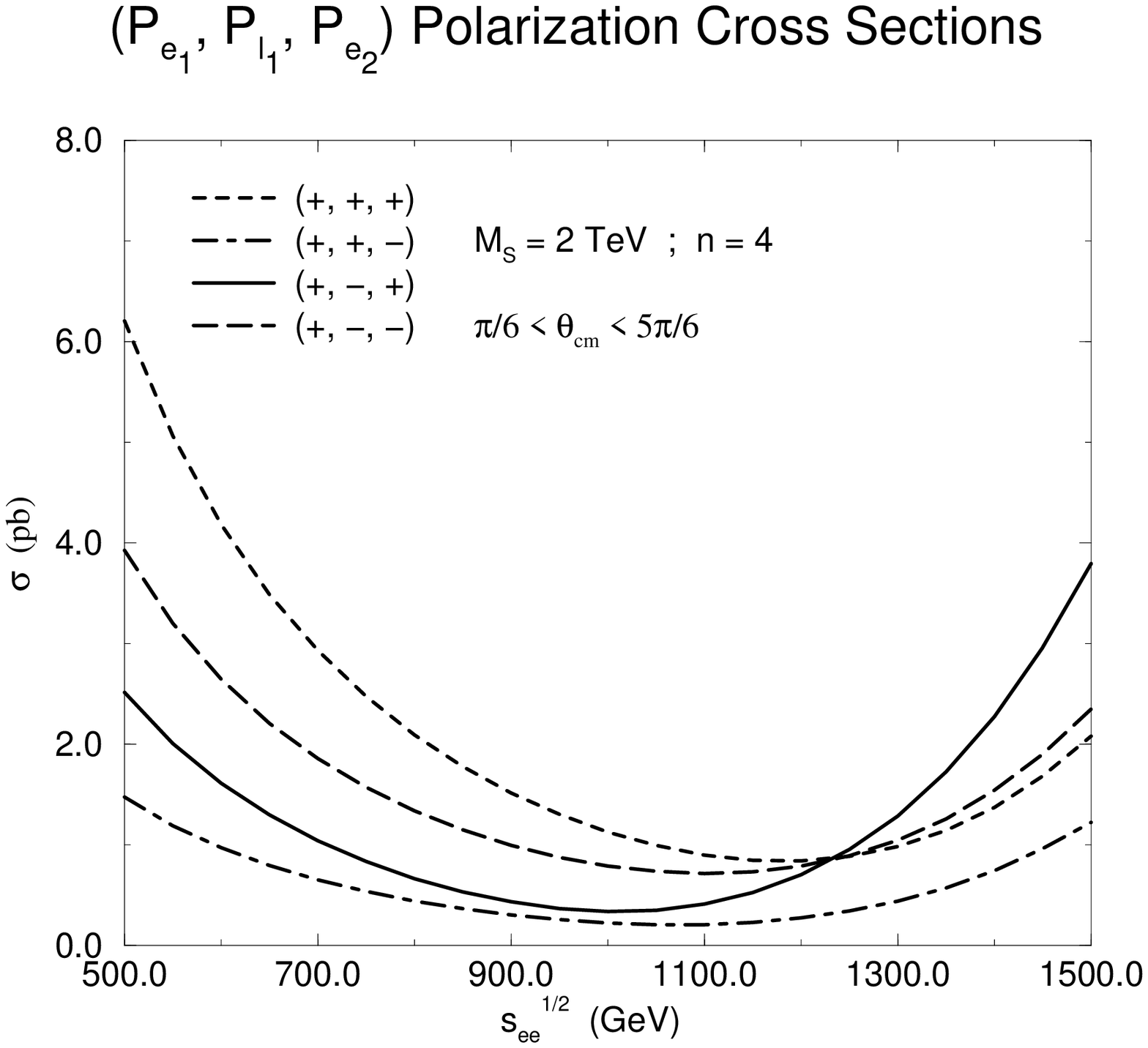}} 
\caption[2]{SM + WSQG cross sections with four independent initial electron and laser beam polarizations.  Here, 
$M_S = 2$ TeV and $n = 4$.} 
\label{polsigs} 
\end{figure}

\begin{figure}[htbp] 
\centerline{\epsfxsize=10truecm \epsfbox{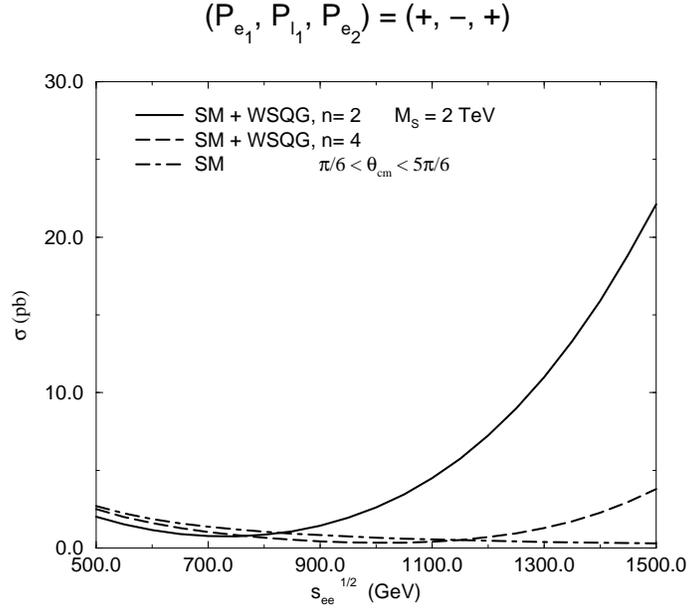}}
\caption[3]{SM + WSQG and SM cross sections for the $(+, -, +)$ polarization.  Here, 
$M_S = 2$ TeV and $n = 2, 4$, for the WSQG contributions.}
\label{24sm}
\end{figure}

In Fig. (\ref{unpcs}), the unpolarized SM cross section and SM $\pm$ WSQG ($w = \pm 1$) cross sections with $M_S = 2$ TeV, $n = 4$,
for $\sqrt{s_{ee}} \in [500, 1500]$ GeV are presented.  The SM + WSQG ($w = +1$) cross sections for $M_S = 2$ TeV, $n = 4$, and
$\sqrt{s_{ee}} \in [500, 1500]$ GeV, with four independent choices of the initial polarization $(P_{e_1}, P_{l_1}, P_{e_2})$ are given in Fig.
(\ref{polsigs}), where the largest high energy polarized cross section is that with the polarization $(+, -, +)$.  In Fig.  (\ref{24sm}), using
the polarization $(+, -, +)$, we compare the SM + WSQG cross sections for $M_S = 2$ TeV and $n = 2, 4$ with the SM cross section, over
the range $\sqrt{s_{ee}} \in [500, 1500]$ GeV.  We see that, in Fig. (\ref{24sm}), the cross section with $n = 2$ is larger than the cross
section with $n = 4$, because of the aforementioned logarithmic enhancement.  Considering the case $w = + 1$ for $n = 4$ in Figs.
(\ref{unpcs}), (\ref{polsigs}), and (\ref{24sm}), we see that the largest departure from the SM result at high energies is obtained by making
use of the $(+, -, +)$ polarization.  

\begin{figure}[htbp]    
\centerline{\epsfxsize=10truecm \epsfbox{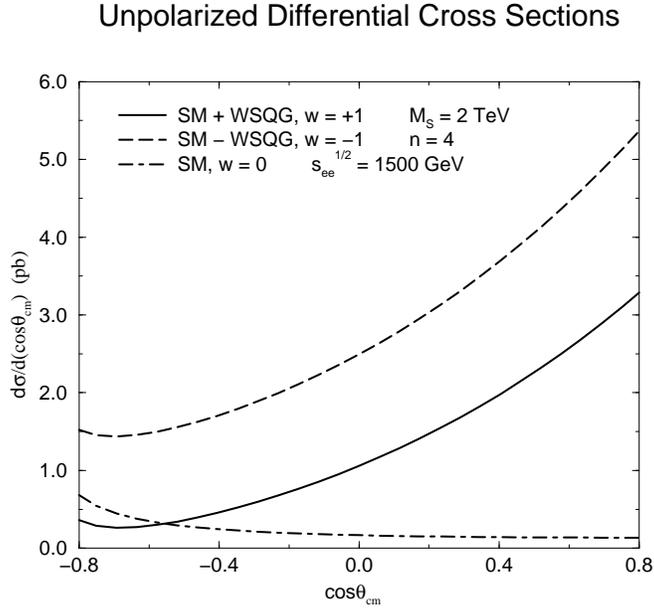}}
\caption[4]
{The unpolarized SM differential cross section and SM $\pm$ WSQG differential cross sections with $M_S = 2$ TeV and $n = 4$, 
at $\sqrt{s_{ee}} = 1500$ GeV.}
\label{unpdcs}
\end{figure}

\begin{figure}[htbp]    
\centerline{\epsfxsize=10truecm \epsfbox{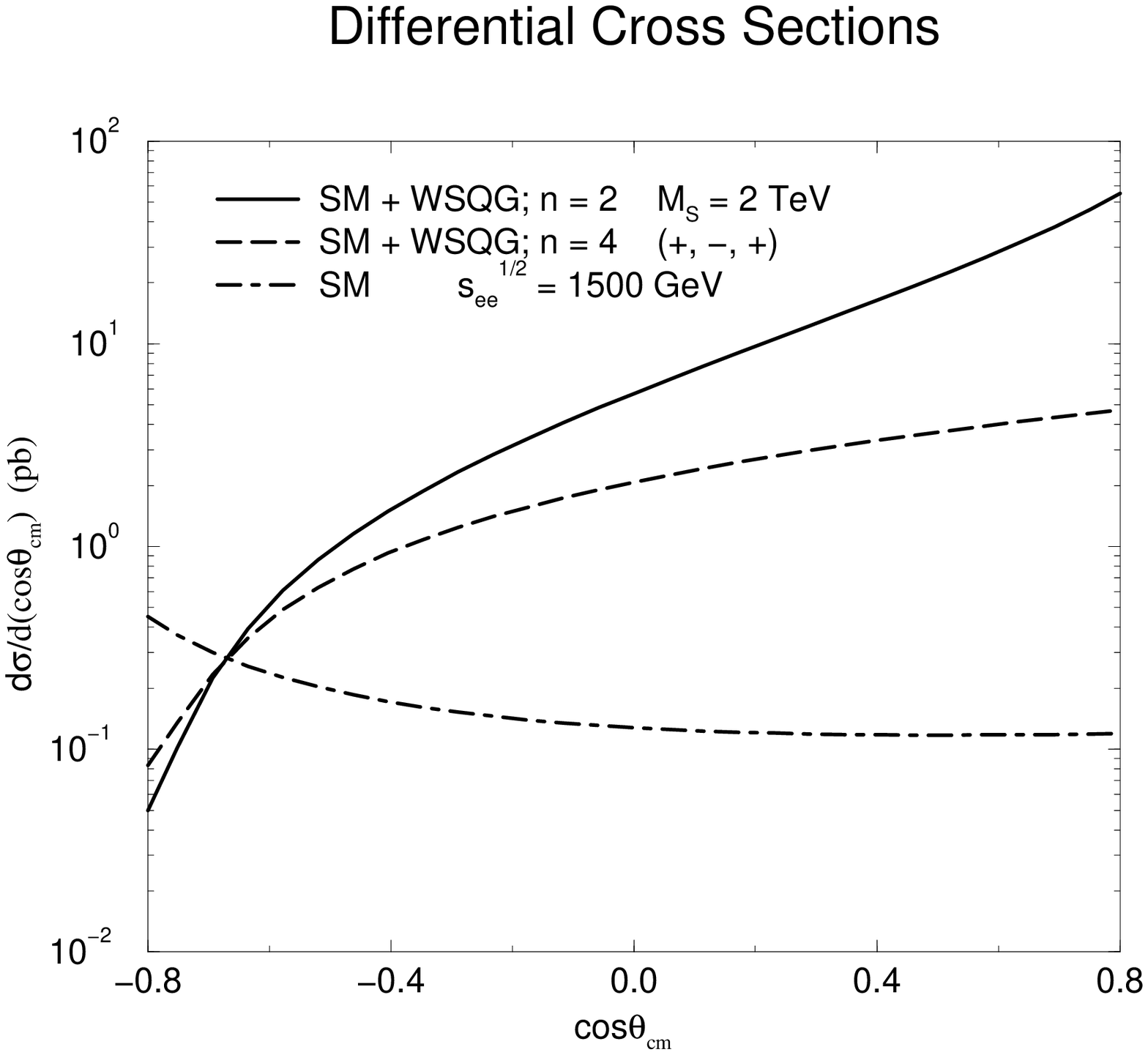}}
\caption[5]
{SM + WSQG and SM differential cross sections at $\sqrt{s_{ee}} = 1500$ GeV for the $(+, -, +)$ polarization.  
Here, $M_S = 2$ TeV and $n = 2, 4$, for the WSQG contributions.}
\label{dcs}
\end{figure}

We present the unpolarized SM differential cross section and SM $\pm$ WSQG differential cross sections with $M_S = 2$ TeV, $n = 4$, at
$\sqrt{s_{ee}} = 1500$ GeV, in Fig. (\ref{unpdcs}).  The differential cross sections with polarization $(+, -, +)$ at $\sqrt{s_{ee}} = 1500$
GeV for SM, and SM + WSQG, with $M_S = 2$ TeV and $n = 2, 4$, are presented in Fig. (\ref{dcs}).  We see that at this value of
$\sqrt{s_{ee}}$, the SM $\pm$ WSQG angular distributions for $\gege$ are very different from the prediction of the SM.  The SM + WSQG
differential cross section with $n = 2$ is enhanced in the forward direction, since $\ln (M_S^2/{\hat t}) \to \infty$ as $\theta_{cm} \to 0$. 
Comparing Figs.  (\ref{unpdcs}) and (\ref{dcs}), we again note that the use of the $(+, -, +)$ polarization results in an enhanced signal.  

\begin{figure}[htbp]    
\centerline{\epsfxsize=10truecm \epsfbox{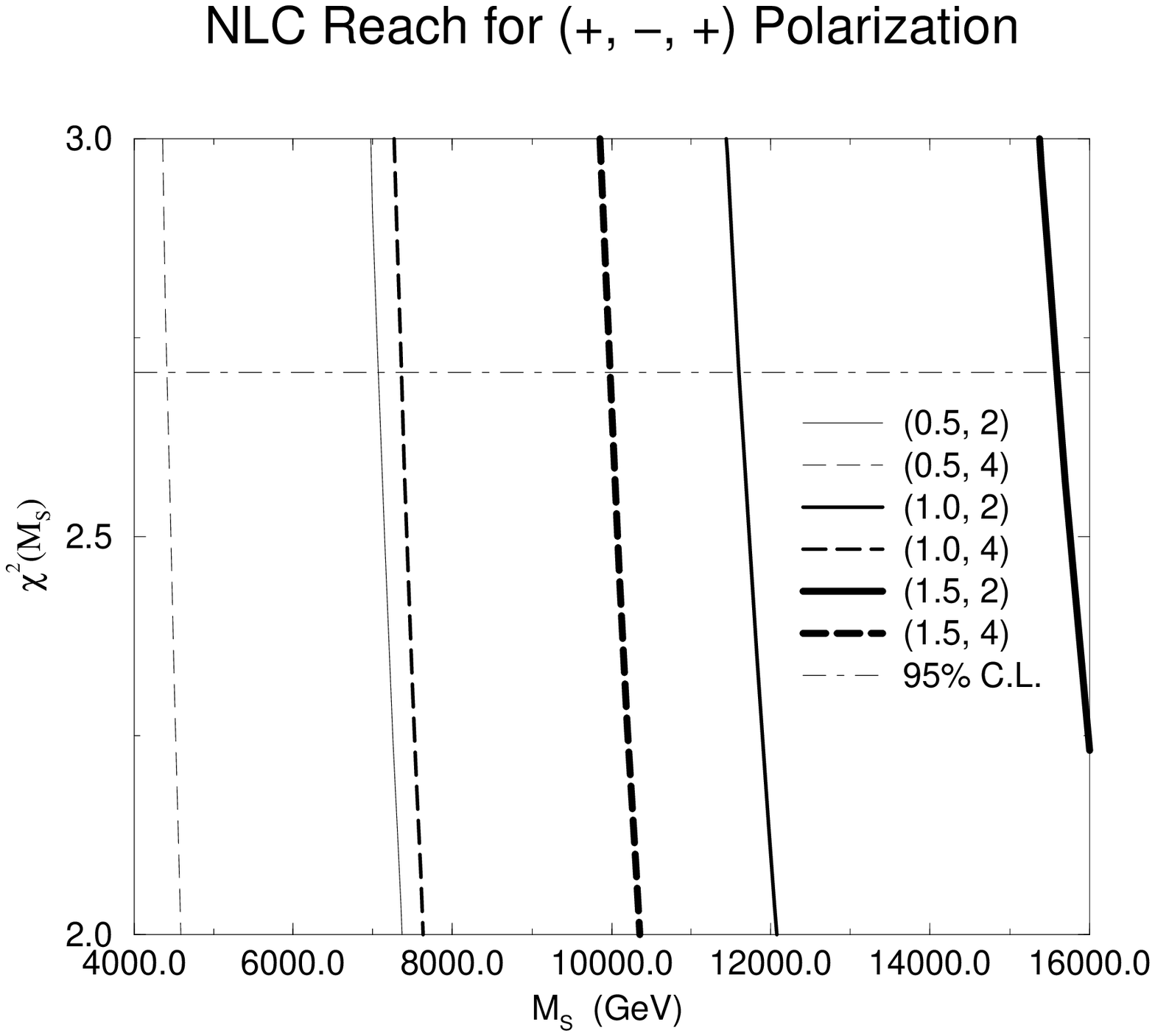}}
\caption[6]{The solid and the dashed lines represent the $\chi^2$ as a function of $M_S$ for the cases 
$n = 2$ and $n = 4$, respectively, at three values of $\sqrt{s_{ee}}$ with polarization $(+, -, +)$.  
The numbers in the parentheses denote the value of 
$\sqrt{s_{ee}}$, in TeV, and $n$, respectively.  The dot-dashed line marks the reach at the $95\%$ confidence level.}
\label{reach}
\end{figure}

To obtain the reach, we have used the $\chi^2 (M_S)$ variable given by 
\begin{equation} \chi^2 (M_S) =
\left(\frac{L}{\sigma_{_{SM}}}\right)\left[\sigma_{_{SM}} - \sigma (M_S)\right]^2, 
\label{chi2} 
\end{equation}
where $L$ is the luminosity, $\sigma_{_{SM}}$ is the SM cross section, and $\sigma (M_S)$ is the SM $\pm$ WSQG cross section as a
function of $M_S$.  We have taken $L = 100$ fb$^{-1}$ per year for our calculations.  To get the reach, we demand $\chi^2 (M_S) \geq
2.706$, corresponding to a one-sided $95\%$ confidence level.  The $M_S$ reach at the NLC with center of mass energies of 500 GeV,
1000 GeV, and 1500 GeV, for the $(+, -, +)$ polarization choice, are shown in Fig. (\ref{reach}).  The smallest reach in Fig. (\ref{reach}) is
about 4 TeV for $n = 4$ and $\sqrt{s_{ee}} = 500$ GeV and the largest reach is about 16 TeV for $n = 2$ and $\sqrt{s_{ee}} = 1500$ GeV. 
Note that the reach for $n = 2$ at $\sqrt{s_{ee}} = 500$ GeV is about 7 TeV or approximately $14 \sqrt{s_{ee}}$.  According to Eq.
(\ref{chi2}), the reach can be improved by increasing the luminosity $L$.  However, we have checked that using $L = 200$ fb$^{-1}$ per
year does not improve the reach significantly.  

\begin{figure}[htbp]    
\centerline{\epsfxsize=10truecm \epsfbox{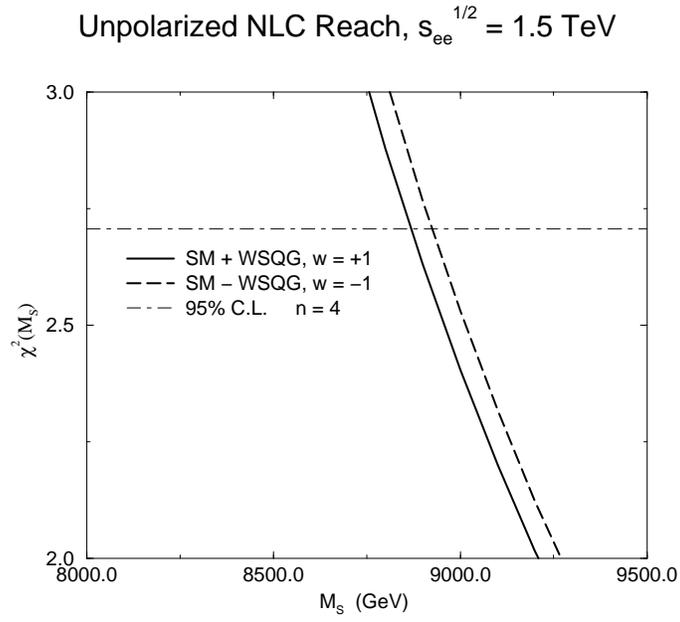}}
\caption[7]{The solid and the dashed lines, corresponding to $w = \pm 1$, respectively, represent the $\chi^2$ as a 
function of $M_S$ for unpolarized beams as a function of $M_S$, with $n = 4$, at $\sqrt{s_{ee}} = 1500$ GeV.  
The dot-dashed line marks the reach at the $95\%$ confidence level.}
\label{unpreach}
\end{figure}

Finally, we mention that the large $M_S$ behavior of the SM $\pm$ WSQG cross section, relevant to the calculation of the reach, can be
inferred from the low energy behavior of the cross sections in Fig. (\ref{unpcs}).  We thus conclude that at large $M_S$, the departure
from the SM result which determines the reach in our calculations, should be roughly the same for $w = \pm 1$.  We also expect the
unpolarized beams to yield a lower reach than that obtained with the $(+, -, +)$ polarization.  These points are demonstrated in Fig. 
(\ref{unpreach}), where we present the unpolarized NLC reach for $n = 4$ and $w = \pm 1$ at $\sqrt{s_{ee}} = 1500$ GeV.  

\section{Concluding Remarks}

In this work, we have shown that the NLC with the photon collider option can be effectively used to constrain theories of weak scale
quantum gravity by measuring the scattering process $\gege$ at TeV energies.  The size of the expected effect depends on the choice of the
electron and laser polarizations.  The results of this paper suggest that studying $\gege$ at the NLC, operating at $\sqrt{s_{ee}} \in [500,
1500]$ GeV and $L = 100$ fb$^{-1}$ per year, can constrain the scale $M_S$ at which quantum gravity becomes important, over the range 4
TeV $\lsim M_S \lsim$ 16 TeV.  This makes $\gege$ one of the most promising discovery channels for weak scale quantum gravity at the
NLC.  

\section*{Acknowledgements}

It is a pleasure to thank J. Hewett, M. Peskin, J. Rathsman, and T. Rizzo  for various comments and conversations.

\appendix
\section*{Appendix A}

In this appendix, we present the expressions that have been used in writing down the gravity amplitude in Eq. (\ref{Mt}).  The Feynman rules
used to obtain Eq. (\ref{Mt}) have been taken form Ref. \cite{Han}, where more details can be found.  The convention used in this paper for
the relation between the fundamental mass scale $M_F$ of gravity and the size $R$ of the $n$ extra dimensions is given by \cite{Han} 
\begin{equation}
\kappa^2 R^n = 16 \pi \, (4 \pi)^{n/2} \, \Gamma (n/2) \, M_F^{-(n + 2)},
\label{RMF}
\end{equation}
where $\kappa = \sqrt {16 \pi G_N}$; $G_N$ is the four dimensional Newton constant and $\Gamma$ represents the
Gamma-function.

The expressions for $C_{\mu \nu, \lambda \sigma}$ and $D_{\mu \nu, \lambda \sigma}(k, p)$,
used in Eq. (\ref{Mt}), are given by \cite{Han}
\begin{equation}
C_{\mu \nu, \lambda \sigma} = \eta_{\mu \lambda} \eta_{\nu \sigma} + \eta_{\mu \sigma} \eta_{\nu \lambda} - \eta_{\mu \nu} \eta_{\lambda \sigma}
\label{Cmnls}
\end{equation}
and
\begin{equation}
D_{\mu \nu, \lambda \sigma}(k, p) = \eta_{\mu \nu} k_\sigma p_\lambda - \left[\eta_{\mu \sigma} k_\nu p_\lambda + 
\eta_{\mu \lambda} k_\sigma p_\nu - \eta_{\lambda \sigma} k_\mu p_\nu + (\mu \leftrightarrow \nu)\right],
\label{Dmnls}
\end{equation}
respectively, where $\eta_{\mu \nu}$ is the Minkowski metric tensor.

\section*{Appendix B} 

In order to calculate various polarization cross sections from Eq. (\ref{diffcs}), we need the photon number density and the average
polarization of the back-scattered $\gamma$ beam.  In this appendix, we provide the expressions for these distribution functions; the
detailed properties of these functions are discussed in Ref. \cite{Ginzburg}.  Let $P_e$ and $P_l$ be the polarizations of the electron beam
and the laser beam, respectively.  We define the function $C(x)$ \cite{Ginzburg} by
\begin{equation}
C(x) \equiv \frac{1}{1 - x} + (1 - x) - 4 r (1 - r) - P_e \, P_l \, r \, z (2 r - 1) (2 - x),
\label{C(x)}
\end{equation}
where $r \equiv x/[z(1 - x)]$.  Then, the photon number density $f(x, P_e, P_l; z)$ is given by
\begin{equation}
f(x, P_e, P_l; z) = \left(\frac{2 \pi \alpha^2}{m_e^2 z \sigma_{_{C}}}\right) C(x), 
\label{f(x)}
\end{equation}
where 
\[
\sigma_{_{C}} = \left(\frac{2 \pi \alpha^2}{m_e^2 z}\right) \left[\left(1 - \frac{4}{z} -\frac{8}{z^2}\right) \ln (z + 1) + \frac{1}{2} + \frac{8}{z} - 
\frac{1}{2 (z + 1)^2}\right]
\]
\begin{equation}
+ P_e \, P_l \left(\frac{2 \pi \alpha^2}{m_e^2 z}\right) \left[\left(1 + \frac{2}{z}\right) \ln (z + 1) - \frac{5}{2} + \frac{1}{z + 1} - \frac{1}{2 (z + 1)^2}\right].   \label{sigC}
\end{equation}
The average helicity $\xi_2(x, P_e, P_l; z)$ is given by
\begin{equation}
\xi_2(x, P_e, P_l; z) = \frac{1}{C(x)}\left\{P_e \, \left[\frac{x}{1 - x} + x (2 r - 1)^2\right] - P_l \, (2r - 1)\left(1 - x + \frac{1}{1 - x}\right)\right\}.
\label{xi2}
\end{equation}

\end{document}